\documentclass{PoS}
\usepackage{lineno}
\usepackage{wrapfig}
\usepackage{graphicx}
\usepackage{csquotes}
\usepackage{subcaption}
\usepackage{url}

\title{Capturing Cosmic Ray Research and Researchers
 with Art}

\ShortTitle{Capturing Cosmic Ray Research and Researchers
 with Art}

\author{
The IceCube Collaboration\footnote{For collaboration list, see PoS(ICRC2019) 1177.}\\
{\itshape \href{http://icecube.wisc.edu/collaboration/authors/icrc19_icecube}{http://icecube.wisc.edu/collaboration/authors/icrc19\_icecube}}\\
E-mail: \email{abduallah@wisc.edu,mdhosale@yorku.ca,ladoni@wisc.edu, james.madsen@uwrf.edu}
}

\abstract{We describe our experiment with an alternate approach to presenting cosmic ray research. The goal was to more widely promote cosmic ray research and attract diverse audiences, especially those from groups that are underrepresented in science or that do not have experience attending science outreach events. The IceCube Neutrino Observatory education and outreach team brought together local teenagers, internationally accomplished artists, science communicators, and scientists to produce an interactive gallery exhibit, {\itshape Messages}, that explores the cosmic ray community and science. The artists collaborated with the scientists and students to create two original installations that will be displayed at the UW--Madison Memorial Union Gallery for six weeks, from mid-June, 2019, through the end of the International Cosmic Ray Conference 2019. {\itshape Event Horizon} by Abdu'Allah with Ladoni features portraits of cosmic ray researchers and high school students who are learning more about the field. This installation will examine the science community as it is and as it could be.  {\itshape Messages from the Horizon} by Hosale with Madsen is inspired by previous immersive works. It combines sound and light to explore what we know and how we know it. 
\\

\vspace{4mm}
{\bfseries Corresponding authors:}
Faisal Abdu'Allah$^{1}$, Mark-David Hosale$^{2}$, Maryam Ladoni$^{1},$\\
\speaker{Jim Madsen}$^{3}$\\

{$^{1}$ \itshape University of Wisconsin--Madison}\\
{$^{2}$ \itshape York University, Toronto, Canada}\\
{$^{3}$ \itshape Univeristy of Wisconsin-River Falls}

}

\FullConference{36th International Cosmic Ray Conference -ICRC2019-\\
		July 24th - August 1st, 2019\\
		Madison, WI, U.S.A.}

\begin{document}

\section{Introduction}\label{intro}
This project uses an alternate approach to examining cosmic ray research and the people who do it. The goal was to more widely promote cosmic ray research and attract diverse audiences, especially those from groups that are underrepresented in science or that do not have experience attending science outreach events. The IceCube Neutrino Observatory education and outreach team brought together local students, internationally accomplished artists, science communicators, and scientists to produce an interactive gallery exhibit that explores the cosmic ray community and science. The artists worked with Madsen and other IceCube scientists and staff to create {\itshape Messages}, which consists of two original installations on display at the UW--Madison Memorial Union Gallery from June 14 through August 2, 2019. {\itshape Event Horizon} (Faisal Abdu'Allah with Maryam Ladoni) features portraits of scientists holding a self-selected meaningful object or artifact. International Cosmic Ray Conference (ICRC) 2019 attendees were invited to bring an artifact or object and have their portrait added to the gallery.  {\itshape Messages from the Horizon} (Mark-David Hosale with Jim Madsen) is an abstracted and collapsed form of the Universe combining sound and light to explore what we know and how we know it.

Planning for the gallery show began more than a year in advance with general discussions between Abdu'Allah, Hosale and Madsen. Hosale and Madsen have been collaborating for about seven years on a variety of art-science projects.  Hosale initiated the partnership when he contacted the IceCube project looking to incorporate data from the IceCube Neutrino Observatory into Quasar 2.0, a sound and light installation that responded to immediate environmental changes while overlaying data from the South Pole.~\cite{Quasar2.0}

\begin{wrapfigure} [11] {r}{.45\textwidth}
\vspace{-.3in}
    \begin{center}
        \includegraphics[width=.4\textwidth]{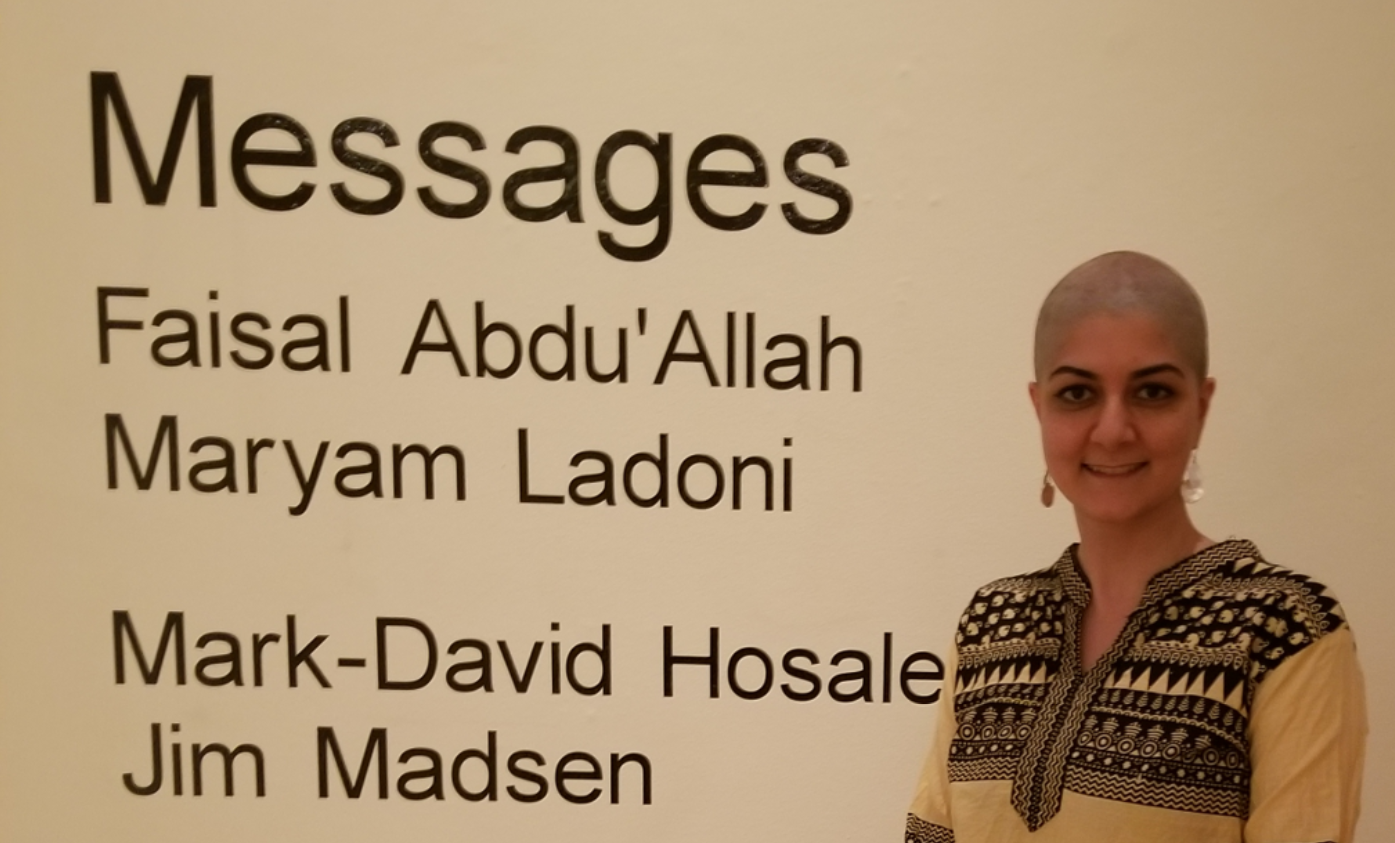}
     \end{center}
\caption{ Ladoni with gallery show graphics. (Photo: Madsen)}
\label{fig:Messages}
\end{wrapfigure}
The connection with Abdu'Allah came about as part of his Squad project~\cite{sqaud} that examines community by selecting and photographing 11 people in each continent. He had completed photo shoots in North America and Europe and had scheduled them in Africa when he contacted IceCube through Madsen to see about getting to Antarctica.  In addition to his incredible art work, Abdu'Allah has also run youth art programs in the Madison Public Library's Teen Bubbler program.~\cite{Bubbler}  

Three facets distinguish {\itshape Messages} from prior IceCube education and outreach efforts. First, by partnering with artists, we were seeking a different way of exploring and communicating cosmic ray science. Abdu'Allah first considered building on his efforts to explore the iconography of a community.  His prior work illustrates the importance of everyday items.  For example, he elevated a barber chair to the equivalent of a throne by gold-plating the metal pieces.  Ultimately, he decided to explore the community of cosmic ray scientists, starting with portraits of three prominent members.  Hosale sought to expand on his world-building, creating his own universe with three persistence-of-vision objects that display stylized forms with accompanying sounds.  A series of four small monitors mimic the process of converting raw information into sensible forms---through simulation, sonification, visualization---and displaying "known" background information.
 
Second, we wanted to reach diverse audiences and communicate in their language.  We partnered with native Spanish-speaking science journalist Angela Posada-Swafford ~\cite{Posada-Swafford}, who translated all the supporting materials for the art exhibit into Spanish and reached out to Spanish news outlets to promote both the exhibit and ICRC 2019.  Third, high school students from underrepresented groups were integrated into content creation for the gallery show.  They were trained by Abdu'Allah to photograph ICRC 2019 attendees whose portraits will be added to the gallery show.  This will provide a real basis for interactions between scientists and students, giving the students a chance to see scientists as people and learn what inspired them to pursue a career in the field of cosmic rays.  It will also hopefully inspire a sense of ownership and a desire to have friends and family visit the gallery to see their work. 
\section{\itshape Event Horizon}
\begin{figure}
\centering
  \begin{subfigure}[b]{0.35\textwidth}
  \centering
    \includegraphics[width=\textwidth]{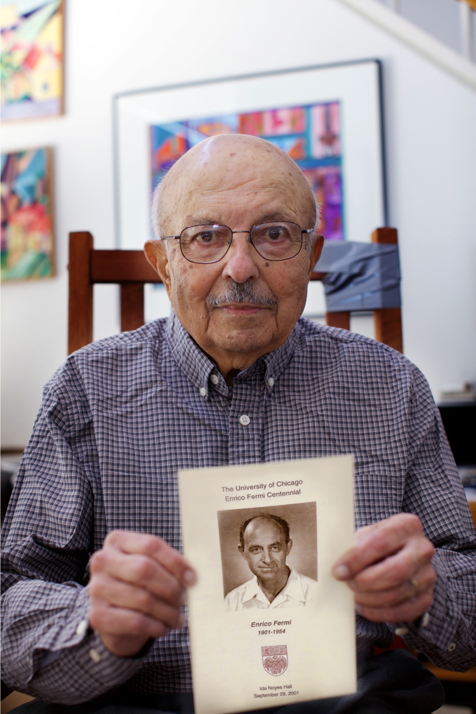}
    \caption{Gaurang Yodh}
    \label{fig:1}
  \end{subfigure}
  \begin{subfigure}[b]{0.35\textwidth}
  \centering
    \includegraphics[width=\textwidth]{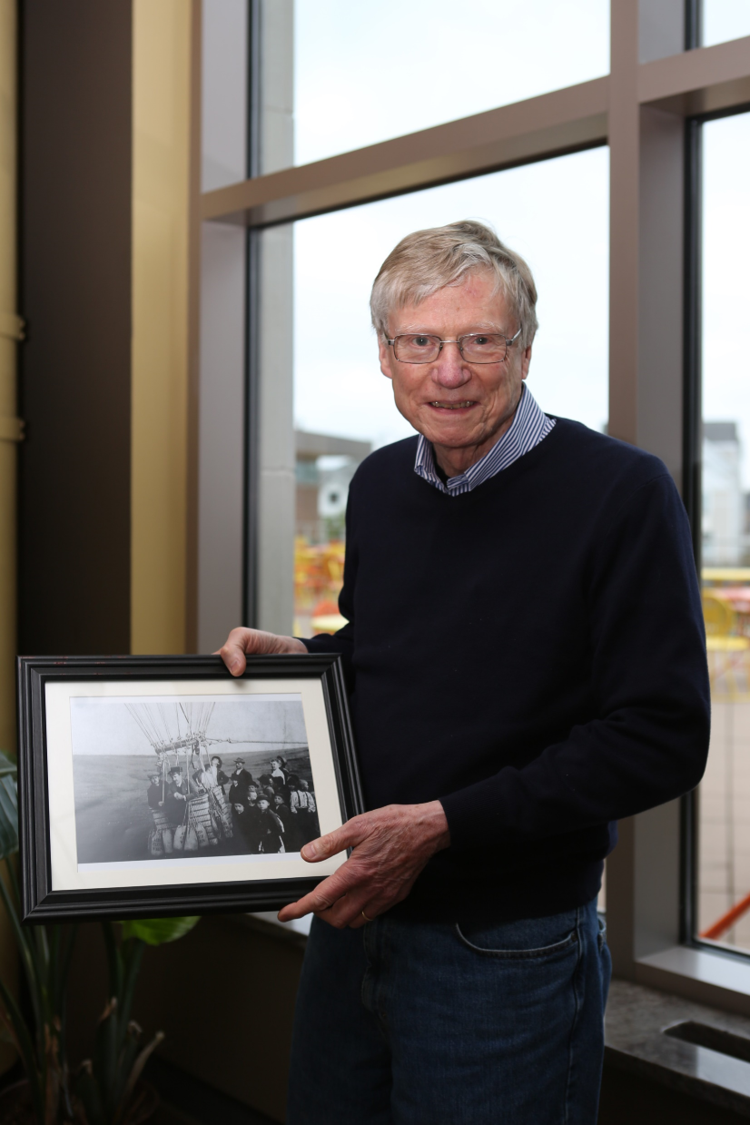}
    \caption{Thomas Gaisser}
    \label{fig:2}
  \end{subfigure}
 \begin{subfigure}[b]{1\textwidth}
  \centering
    \includegraphics[width=.7\textwidth]{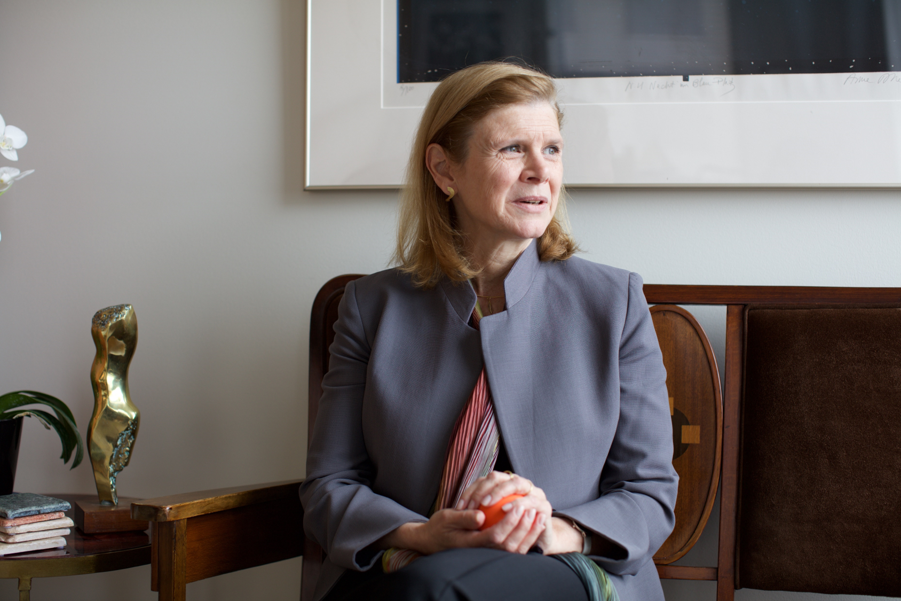}
    \caption{Angela Olinto}
    \label{fig:3}
  \end{subfigure}
\caption{ Portraits of cosmic ray scientists---Gaisser by Ladoni, Olinto and Yodh by Abdu'Allah.}
\end{figure}
Fasial Abdu'Allah worked with Maryam Ladoni on this installation, which consists of photographic prints on canvas; see Figure 2.  Abdu'Allah traveled to Chicago, IL, to photograph Professor Angela Olinto in her office and to Irvine, CA, to get the portrait of Professor Gaurang Yodh in his home.  Ladoni took the picture of Professor Thomas Gaisser in Madison, WI, on the UW--Madison campus while Gaisser was attending the spring 2019 IceCube Collaboration meeting.  Gaurang Yodh's portrait is especially meaningful as he died about two months later, on June 3, 2019, at the age of 90.  Each individual was instructed to select an artifact or object to hold for their portrait as a way to reveal something about themselves.  Yodh chose a picture of Enrico Fermi, his thesis advisor. Gaisser is holding the famous picture of Victor Hess in his balloon after taking radiation measurements as a function of height. Gaisser visited the site in Germany where the balloon landed on the 100-year anniversary of the historic event. Olinto is holding a ball that she tosses while thinking to help her focus.    

\subsection {Artist Statement}
{\itshape Messages} poetically describes a place without boundaries---where artists find inspiration. Artisans have always been the vanguards of protest and shapers of social consciousness. {\itshape Event Horizon} is a public art collaboration that seamlessly illustrates the relationship between art and the sciences. Three eminent professors (Gaurang Yodh, Thomas Gaisser, and  Angela Olinto) were asked to think of an object or artifact that has been meaninful in their science careers and were documented holding the object in the palm of their hands in a generous act of sharing. Throughout the course of the show, additional images will populate the gallery from ICRC attendees and other individuals holding their objects of inspiration.

\subsection {Artists' Bios}
Faisal Abdu'Allah graduated from the Royal College of Art [Ph.D. dissertation "Mirror to my Thoughts" under the supervision of the artist Gavin Turk (YBA)] in 2012. He is cited in over 50 publications, has lectured on four continents, and was the first Afro-British visiting Professor of Art to exhibit at Stanford University. He has exhibited at Tate Modern, the Studio Museum in Harlem, the Serpentine Gallery, the 55th Venice Biennale, and the Royal Academy. His works are also in the collections of Tate Britain, The Victoria \& Albert Museum, the California African American Museum (CAAM), the National Maritime Museum, the British Arts Council, and the Chazen Museum of Art. Abdu'Allah was awarded first prize at Tallinn Print Triennial, the Joan Mitchell Painters \& Sculptors Grant, the Mayor of London Arts Award, and the Romnes Faculty Award. His work will be featured in \textbf {Get Up Stand Up} at Somerset House, London (2019), and the Foto-Fest Biennial 2020 central exhibition, \textbf {African Cosmologies}, Houston, TX. He is represented by Magnolia Editions, USA, and Autograph (ABP), UK, and maintains studios in London and Madison and currently resides in the US.

Maryam Ladoni is a photo-based artist living in the US. She received an MA degree in photography from the University of Art in Iran. Currently, she is an MFA candidate (2020) in photography at the University of Wisconsin--Madison. She also works as a teaching assistant in the Digital Print Center and as a photographer in the Division of the Art at UW--Madison.

\section{\itshape Messages from the Horizon}
Mark-David Hosale worked with Jim Madsen on this installation, which consists of kinetic persistence-of-vision displays combined with embedded computers with screens and custom electronics; see Figure 3.   
\begin{figure}
\centering
  \begin{subfigure}[b]{1\textwidth}
  \centering
    \includegraphics[width=.7\textwidth]{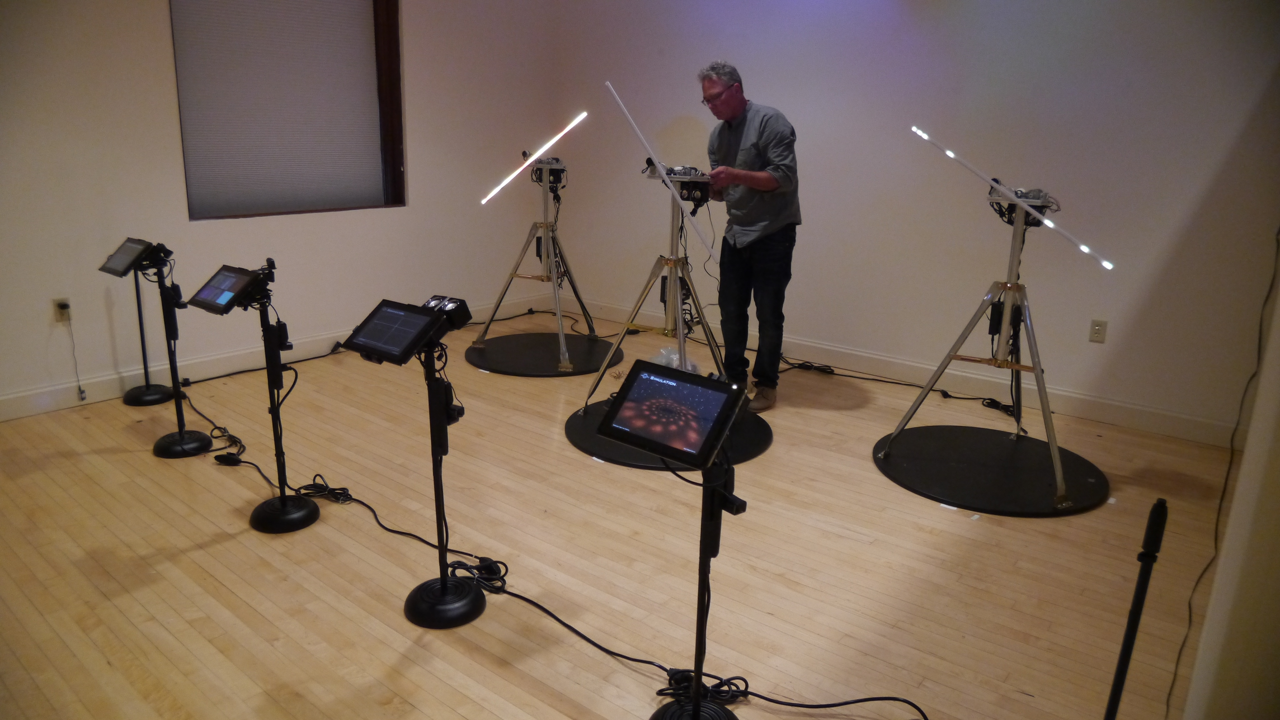}
    \caption{Madsen working on install.}
    \label{fig:4}
  \end{subfigure}
 \begin{subfigure}[b]{1\textwidth}
  \centering
    \includegraphics[width=.7\textwidth]{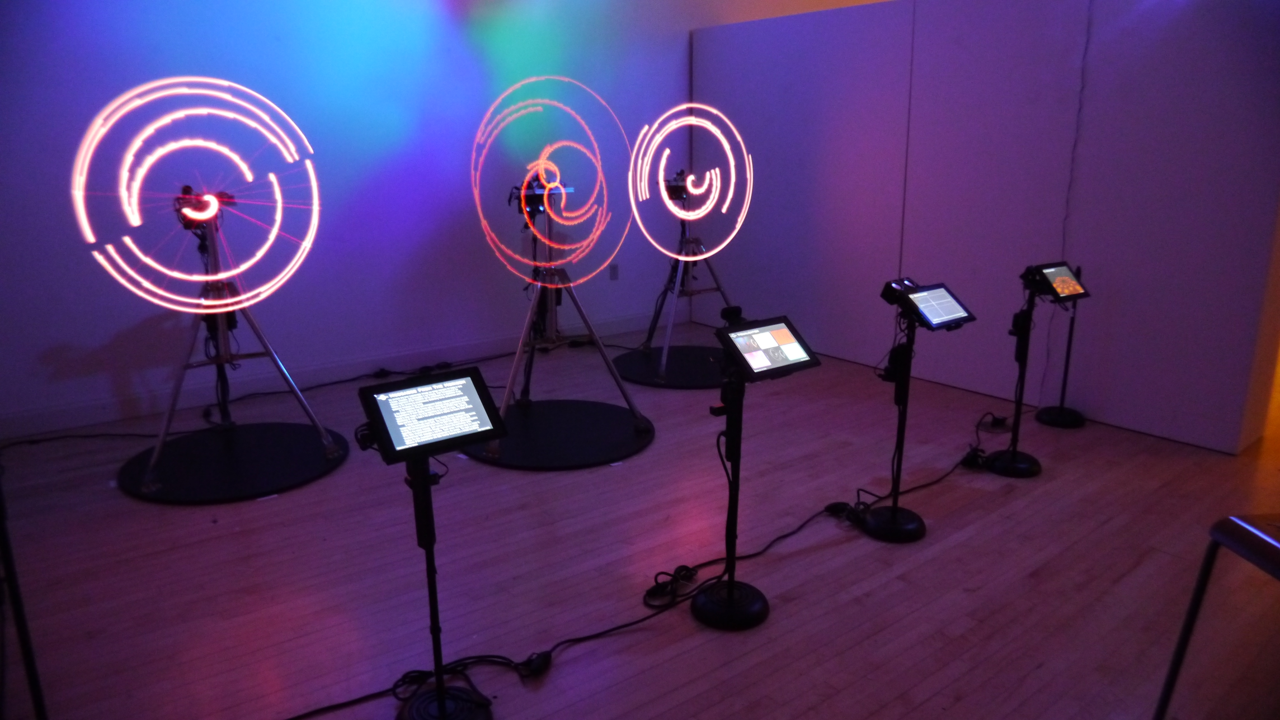}
    \caption{\itshape Messages from the Horizon}
    \label{fig:5}
  \end{subfigure}
\caption{Photos by Hosale showing installation and the exhibit while in operation.}
\end{figure}

\subsection {Artist Statement}
{\itshape Messages from the Horizon} is an abstracted and collapsed form of the Universe within the space of a small room. It is an expression of the process of astronomical observation, knowledge acquisition, and the human condition in the Universe. Most of what we wish to learn about our Universe falls outside of what we are able to perceive through our senses. Ultimately, all information gathered from our instruments must be scaled and transformed so that it can fall within the range of our perception to be understood. In the case of astronomical observation, we do this by slicing the electromagnetic spectrum and comparing significant simultaneous events in our Universe in energy, location, and time. The relation of the totality of what is present in the Universe and what is perceivable through our senses highlights the edges of what is known, knowable, and unknowable in our Universe. From this perspective, the abstract experience of this installation can be understood as an exploration of our relationship to the Universe by simultaneously celebrating human achievement and the existential limitations of our being.

In the installation, three semitransparent displays represent black holes located in the distant reaches of our Universe using a technique known as persistence of vision. The persistence-of-vision display consists of a single string of LEDs spun at a high speed to draw images in the air by imprinting on our eyes as it spins. The effect of persistence of vision is used in this exhibit to express the elusive nature of black holes and their study.

In nature, black holes are sources of high-energy activity that produce messengers that eventually travel to Earth. The virtual black holes represent these sources, and the array of four monitors collect and assemble their cosmic message. Each monitor offers a different perspective of the same virtual phenomenon through visualizations and sonifications of the simulated black hole messages in real time. Touch screen interfaces allow the viewer to explore the data visualizations being presented on the screens. The result is a partial reflection of the phenomenon with each monitor showing only a limited and altered impression of the event that has occurred.

\subsection {Artists' Bios}
Mark-David Hosale is a computational artist and composer, and he is an associate professor and chair of computational arts in the School of the Arts, Media, Performance, and Design, York University, Toronto, Canada. His solo and collaborative work has been exhibited internationally at such venues as the SIGGRAPH Art Gallery (2005); International Symposium on Electronic Art (ISEA 2006); BlikOpener Festival, Delft, The Netherlands (2010); the Dutch Electronic Art Festival (DEAF2012); Biennale of Sydney (2012); Toronto's Nuit Blanche (2012); Art Souterrain, Montreal (2013); and a collateral event at the Venice Biennale (2015), among others. 

His interdisciplinary practice is often built on collaborations with architects, scientists, and other artists in the field of computational arts. Hosale's work explores the boundaries between the virtual and the physical world that results in a disparate practice that spans from performance (music and theatre) to public and gallery-based art. The connecting tissue in his work is an interest in knowing. How do we come to know something? How do we know we know? And, how do we express what we know to each other? Through immersive art we are able to create new experiences that saturate the individual, expressing concepts that are beyond language and only genuinely knowable through the senses. 

Jim Madsen is an associate director of the IceCube Neutrino Observatory responsible for education and outreach efforts and a professor of physics at the University of Wisconsin-River  Falls.  He enjoys crossing boundaries, bringing together people with different ways of exploring and communicating. The intriguing combination of the extremes associated with the IceCube project---the South Pole location, ghost-like cosmic messenger neutrinos, and cataclysmic astrophysical phenomena---are awe inspiring.  The artists connections are motivated by the belief that a deeper understanding of something is gained by viewing it from multiple vantage points. 

\section{Summary}

To capitalize on the energy and excitement of hosting ICRC 2019 at UW--Madison, the IceCube Neutrino Observatory education and outreach team brought together local teenagers, internationally accomplished artists, science communicators, and scientists to produce an interactive gallery exhibit that explores the cosmic ray community and their research. Just as scientists must look in new ways to better understand their subject matter, we must try new approaches to connect with diverse groups and reach different audiences. ICRC 2019 is a unique opportunity to increase science, technology, engineering, and mathematics engagement and a chance to promote cosmic ray research beyond the research community. We leveraged existing partnerships to reach a diverse audience by employing three strategies. First, we explored both the science and the people who do it from the perspective of two accomplished, international artists. Second, we involved teens from underrepresented groups to develop ownership, and engage in a meaningful mentoring experience. Third, we worked with a high-profile bilingual science communicator to provide all content in English and Spanish and promote the experience both locally and nationally.\\
\bigskip 

\textbf{Acknowledgements}

This work was supported by funds from the Wisconsin IceCube Particle Astrophysics Center (WIPAC) and the Local Organizing Committee of ICRC 2019.  Special thanks goes to Faisal Abdu'Allah, Marayam Ladoni and Mark-David Hosale for helping us see in new ways.  We greatly appreciate the willingness of Professors Gaissser, Olinto and Yodh to participate in this project. We want to express our condolences to the family and friends of Professor Gaurang Yodh. We thank the Wisconsin Union Directorate for the opportunity to use the Memorial Union Gallery and Lily Miller, Sophie Plzak, and Tony Wise for their help. Finally, special thanks to Angela Posada-Swafford and Jazmine Zuniga Paiz for their Spanish translations and promotional material, and to the IceCube education and outreach team at WIAPC, Ellen Bechtol, Jean DeMerit, Madeleine O'Keefe, and Lindsey Steffes.



\bibliographystyle{ICRC}
\bibliography{references}

%

\end{document}